# Passivation-induced physicochemical alterations of the native surface oxide film on 316L austenitic stainless steel

Zuocheng Wang, Francesco Di-Franco,[1] Antoine Seyeux, Sandrine Zanna, Vincent Maurice,[2] Philippe Marcus[3]

*PSL Research University, CNRS - Chimie ParisTech, Institut de Recherche de Chimie Paris, Research Group Physical Chemistry of Surfaces, Paris, France*

## Abstract

Time of Flight Secondary Ion Mass Spectroscopy, X-Ray Photoelectron Spectroscopy, *in situ* Photo-Current Spectroscopy and electrochemical analysis were combined to characterize the physicochemical alterations induced by electrochemical passivation of the surface oxide film providing corrosion resistance to 316L stainless steel. The as-prepared surface is covered by a ~2 nm thick, mixed (Cr(III)-Fe(III)) and bi-layered hydroxylated oxide. The inner layer is highly enriched in Cr(III) and the outer layer less so. Molybdenum is concentrated, mostly as Mo(VI), in the outer layer. Nickel is only present at trace level. These inner and outer layers have band gap values of 3.0 and 2.6-2.7 eV, respectively, and the oxide film would behave as an insulator. Electrochemical passivation in sulfuric acid solution causes the preferential dissolution of Fe(III) resulting in the thickness decrease of the outer layer and its increased enrichment in Cr(III) and Mo(IV-VI). The further Cr(III) enrichment of the inner layer causes loss of photoactivity and improved corrosion protection with the anodic shift of the corrosion potential and the increase of the polarization resistance by a factor of ~4. Aging in the passive state promotes the Cr enrichment in the inner barrier layer of the passive film.

---

[1] Present address : Electrochemical Materials Science Laboratory, DICAM, Universita degli Studi di Palermo, Palermo, Italy
[2] Corresponding author : vincent.maurice@chimieparistech.psl.eu
[3] Corresponding author : philippe.marcus@chimieparistech.psl.eu

# 1. Introduction

Stainless steels (SS) are widely used thanks to their excellent mechanical properties and corrosion resistance. Surface analytical studies, including X-Ray Photoelectron Spectroscopy (XPS), performed on ferritic Fe-Cr(-Mo) [1-19] and austenitic Fe-Cr-Ni(-Mo) [20-40] SS substrates have shown that a continuous and protective surface oxide/hydroxide layer, the passive film, markedly enriched in Cr(III) and only a few nanometers thick when formed at ambient temperature, provides the corrosion resistance. The Cr enrichment in the passive film, which is the key factor for the corrosion resistance, is strong in acid aqueous environment because of the competitive dissolution of the iron and chromium oxide species and comparatively small dissolution rate of Cr(III) oxide. In alkaline aqueous solutions, the lower solubility of Fe(II)/Fe(III) oxides mitigates the Cr enrichment. For Ni-bearing SS, there is no or very little Ni(II) in the passive films and the metallic alloy region underneath the oxide is enriched with Ni(0) [20-22,27-30]. Mo-bearing SS, including austenitic AISI 316L, have better corrosion resistance in chloride-containing environments, where passive film breakdown can be followed by the initiation of localized corrosion by pitting. Mo(IV) or Mo(VI) oxide species enter the passive film composition at a few at% level without markedly altering the thickness [10,21,22,27,30,32,35,37]. Molybdenum has been proposed to mitigate passive film breakdown or to promote passive film repair.

The Cr(III) enrichment may not be homogeneous in the passive film, as suggested by recent studies performed at the nanometer scale [27,41], and the heterogeneities may cause the local failure of the corrosion resistance and of the initiation of localized corrosion [42]. These heterogeneities of Cr enrichment would find their origin in the mechanisms of pre-passivation leading to the initial formation, most often in air, of the native oxide and in the subsequent modifications of the oxide film induced by immersion in solution. The better understanding of the mechanisms governing the Cr enrichment requires to thoroughly investigate both the initial stages of oxidation leading to pre-passivation of the SS surface [43] and the modifications brought by electrochemical passivation of the native oxide-covered SS surface.

Here we report on the properties of the native oxide layer formed on a polycrystalline austenitic 316L SS surface in air and modified by electrochemical passivation in acid solution. Time-of-Flight Secondary Ion Mass Spectroscopy (ToF-SIMS) and XPS surface analysis were used to characterize the layered structure, thickness and composition of the oxide film and their modifications. Photocurrent Spectroscopy (PCS), applied *in situ* and combined with electrochemical analysis [44-47], was used to obtain direct information on the electronic properties of the thin photo-conducting surface films. Linear Scan Voltammetry and Electrochemical Impedance Spectroscopy were applied to discuss the corrosion resistance.

## 2. Experimental

Polycrystalline AISI 316L austenitic SS samples were used. The bulk wt% composition of the main alloying elements was Fe–19Cr–13Ni–2.7Mo (Fe–20Cr–12Ni–1.6Mo at%). Surface preparation was performed by mechanical polishing first with emery paper of successive 1200 and 2400 grades and then with diamond suspensions of successive 6, 3, 1 and 0.25 µm grades. Cleaning and rinsing was performed after each polishing step in successive ultrasonicated baths of acetone, ethanol and Millipore® water (resistivity > 18 MΩ cm). Filtered compressed air was used for drying. The $R_a$ roughness of the as-polished surfaces was 2.4 nm measured by AFM on $10 \times 10$ µm$^2$ areas. The native oxide film was analyzed after short time (5-10 min) exposure in ambient air after surface preparation. The passive oxide film was formed by electrochemical passivation in a 3-electrode cell. The potential was stepped from open circuit value ($U_{OC}$ = -0.23 V vs SCE) to $U_{Pass}$ = 0.3 V vs SCE and maintained to this value for 1 or 20 hours. The selected value corresponded to the minimum of passive current in 0.05 M $H_2SO_4$ aqueous solution as determined from polarization curves. The electrolyte was prepared from ultrapure chemicals (VWR®) and Millipore® water. No cathodic pre-treatment was performed in order to avoid any reduction-induced alteration of the initial native oxide film prior to passivation. The $R_a$ roughness of the passivated surfaces was 2.1 nm measured by AFM on $10 \times 10$ µm$^2$ areas.

Depth profile elemental analysis of the oxide-covered surfaces was performed using a ToF-SIMS 5 spectrometer (Ion ToF – Munster, Germany) operated at about $10^{-9}$ mbar. The depth profile analysis interlaced topmost surface analysis in static SIMS conditions using a pulsed 25 keV Bi$^+$ primary ion source delivering 1.2 pA current over a $100 \times 100$ µm$^2$ area with sputtering using a 1 keV Cs$^+$ sputter beam giving a 32 nA target current over a $300 \times 300$ µm$^2$ area. Analysis was centered inside the eroded crater to avoid edge effects. The profiles were recorded with negative secondary ions that have higher yield for oxide matrices than for metallic matrices. The Ion-Spec software was used for data acquisition and processing.

Surface chemical analysis was performed by XPS with a Thermo Electron ESCALAB 250 spectrometer operating at about $10^{-9}$ mbar. The X-ray source was an AlK$_\alpha$ monochromatized radiation (hν = 1486.6 eV). Survey spectra were recorded with a pass energy of 100 eV at a step size of 1 eV. High resolution spectra of the Fe 2p, Cr 2p, Ni 2p, Mo 3d, O 1s, and C 1s core level regions were recorded with a pass energy of 20 eV at a step size of 0.1 eV. The take-off angles of the analyzed photoelectrons were 40° and 90°. The binding energies (BE) were calibrated by setting the C 1s signal corresponding to olefinic bonds (–CH$_2$–CH$_2$–) at 285.0 eV. Reconstruction of the spectra (curve fitting) was performed with CasaXPS, using a Shirley type background. Asymmetry was taken into account for the metallic components (Cr$^0$, Fe$^0$, Mo$^0$, Ni$^0$) by using the peak shape of LA(α, β, m) [48]. Lorentzian/Gaussian (70%/30%) peak shapes and a broader envelope was used to account for the multiplet splitting of the oxide components (Cr$^{III}$, Fe$^{III}$, Mo$^{IV}$, Mo$^{VI}$) [48,49].

Photoelectrochemical and electrochemical analysis was performed using a Zennium electrochemical workstation (ZAHNER, Germany) operated with the Thales software. The measurements were performed in 0.1 M Na$_2$B$_4$O$_7$ electrolyte (di-Sodium tetraborate, pH ~9.5) prepared from pure chemicals (R.P.NORMAPUR®) and Millipore® water using a standard 3-electrode cell with a Pt spiral as the counter electrode and an Ag/AgCl (in 3.5 M KCl) electrode as reference electrode. This buffer solution was selected in order to provide stability to the oxide films during measurements and thus to minimize their alteration. The photocurrent spectra were recorded in the wavelength range from 310 to 700 nm with the Zahner Zennium CIMPS system (TLS-03) at a 3.16 Hz frequency. The electrochemical impedance spectra were generated by applying a sinusoidal signal of 10 mV amplitude over the frequency range 0.1 Hz–100 kHz. The spectra were analyzed with the Zview software

## 3. Results and discussion

### 3.1 Surface analysis

#### 3.1.1 Native oxide film

The ToF-SIMS negative ion depth profiles of the 316L stainless steel sample covered by the native oxide film are shown in Fig. 1. In Fig. 1a, the intensity profiles of secondary ions characteristic for the oxide film ($^{18}$O$^-$, $^{18}$OH$^-$, CrO$_2^-$, FeO$_2^-$, NiO$_2^-$ and MoO$_2^-$) and the substrate (Cr$_2^-$, Fe$_2^-$ and Ni$_2^-$) are plotted in logarithmic scale versus sputtering time. As proposed previously [50], the profile of the Ni$_2^-$ ions is used to define the position of the "modified alloy" region between the oxide film and the metallic bulk substrate regions, where Ni is found enriched as observed previously on austenitic and duplex stainless steels [20-22,27-30,50]. This region is measured from 20 to 48 s of sputtering time, meaning that the oxide film region is from 0 to 20 s and the metallic substrate region after 48 s. The sputtering rate can be estimated to 0.1 nm s$^{-1}$ based on the thickness of the oxide film of 2 nm determined by XPS as discussed below. In the surface oxide region, the most predominant profiles of the oxidized metals are those of the $CrO_2^-$, $FeO_2^-$ and to a lesser extent $MoO_2^-$ ions while that of the $NiO_2^-$ ions has much lower intensity, suggesting that only traces of oxidized nickel are present in the native oxide film as supported by the XPS analysis discussed below and in agreement with previous data [20-22,27-30].

Fig. 1b and Fig. 1c compare the profiles of the secondary ions that are characteristic for the oxide species predominantly present in the surface film. Fig. 1b shows that the profiles of $FeO_2^-$ and $CrO_2^-$ ions peak at different positions, at 6 and 10 s, respectively. This is consistent with the native oxide film having a bilayer structure with iron and chromium oxides more concentrated in the outer and inner layers, respectively, as previously reported for oxide films on stainless steels [7,17,27,29,46]. The interface between outer and inner layer can be positioned at 8 s which is the median sputtering time position of the two intensity maxima. In Fig. 1c, the $FeO_2^-$ and $MoO_2^-$ ions profiles peak at the same position (6 s), showing very similar in-depth distributions of the iron and molybdenum oxide species concentrated in the outer layer of the oxide film. In order to confirm the partition of the concentration of the two major elements in the oxide film, the intensity ratio of the $CrO_2^-$ to $FeO_2^-$ ions has been plotted *vs* sputtering time in Fig. 1d. According to this graph, the Cr/Fe ratio in the oxide increases in the outer part of the film. In the inner layer of the oxide film, the Cr/Fe ratio increases until 15 s and then becomes constant.

Fig. 2 shows the angle-resolved XPS Cr $2p_{3/2}$, Fe $2p_{3/2}$, Mo 3d, Ni $2p_{3/2}$ and O 1s core level spectra recorded at 90 and 40° take-off angles. For the Cr $2p_{3/2}$, Fe $2p_{3/2}$ and Mo 3d core levels, the intensity ratio of the oxide peaks at higher BE (see below) to the metal peaks at lower BE (see below) increases with decreasing take-off angles, confirming that the oxide film lies above the metallic alloy substrate. The Ni $2p_{3/2}$ core level spectra show no oxide peaks at higher BE and an intensity decrease at lower take-off angle of the metal peaks. This is because Ni, only present in the metallic alloy, is increasingly attenuated by the surface oxide film. For the O 1s core level, the intensity ratio of the higher BE peak associated to OH$^-$ ligands (see below) to the lower BE peak associated to O$^{2-}$ ligands (see below) is slightly higher at 90° than at 40°. This indicates some stratification in the depth distribution of the oxide/hydroxide ligands according to the bilayer structure of the oxide film with the hydroxide species more concentrated in the outer layer and the oxide species in the inner layer. This stratification was repeatedly observed and more marked in the passive films.

Fig. 3 shows the reconstruction of the XPS Cr $2p_{3/2}$, Fe $2p_{3/2}$, Mo 3d and O 1s core level spectra for the native oxide-covered sample. The BE values, Full Widths at Half-Maximum (FWHM) values and relative intensities of the component peaks obtained by curve fitting are compiled in Table. 1. The Cr $2p_{3/2}$ spectrum (Fig. 3a) was reconstructed using one peak (Cr1) at 574.1 eV, assigned to metallic $Cr^0$ in the substrate and a series of five peaks (Cr2-Cr6), assigned to $Cr^{III}$ oxide in the surface oxide layer, as proposed previously [48,49]. It was verified that such a series of five peaks could fit the spectrum for an anhydrous thermal Cr oxide film grown by exposing a Cr substrate to gaseous oxygen in the XPS preparation chamber. In the present case, one additional peak (Cr7) at 577.2 eV, was necessary to optimize the fit with the experimental envelope. Given the lower relative intensity of this Cr7 component, it was deliberately chosen to use a single but wider peak rather than a series of five narrower peaks as done for the oxide component. This broad and relatively weak Cr7 component was assigned to $Cr^{III}$ hydroxide in the oxide film [2,16,20], No additional $Cr^{VI}$ component at a BE of 579.5 eV [48] was needed. Similarly, in the Fe $2p_{3/2}$ spectrum (Fig. 3b), the peak at 706.8 eV (Fe1) relates to metallic $Fe^0$ in the substrate and the series of five peaks (Fe2-Fe6) to $Fe^{III}$ in the surface oxide layer [7,23,34,48,51,52]. The additional peak needed at 711.9 eV (Fe7) is assigned to $Fe^{III}$ hydroxide in the oxide film. No series of five peaks associated to $Fe^{II}$ in the surface oxide layer was needed for reconstruction [48]. In the Mo $3d_{5/2-3/2}$ core level spectrum (Fig. 3c), the 5/2-3/2 doublet (Mo1/Mo1') of the metallic $Mo^0$ component is positioned at 227.5-230.6 eV [7,36]. The 5/2-3/2 doublets at 229.5-232.6 eV (Mo2/Mo2') and 232.6-235.7 eV (Mo3/Mo3') are assigned to $Mo^{IV}$ and $Mo^{VI}$ in the surface oxide layer, respectively [22,27,31,33,36,53]. The intensity ratio of the $Mo^{VI}$ to $Mo^{IV}$ doublets is ~8, indicating that $Mo^{VI}$ species are mostly present. The O 1s spectrum (Fig. 3d) was reconstructed with three components at 530.3 (O1), 531.6 (O2) and 532.9 (O3) eV assigned to the oxide ($O^{2-}$), hydroxide ($OH^-$) and water ($H_2O$) ligands in the oxide film [11,34]. The broader O3 component may be explained by surface contamination by species including double bonded oxygen. The $OH^-/O^{2-}$ intensity ratio is 0.9 for the native oxide. The Ni $2p_{3/2}$ spectrum (Fig. 2e) could be reconstructed with a single peak at 852.9 eV BE corresponding to metallic $Ni^0$ in the substrate [24], evidencing that the Ni oxide species, measured in the oxide film region by ToF-SIMS, were at trace level below the detection limit of XPS (~0.5 at%).

Based on the partition of the oxide film evidenced by the ToF-SIMS data and the depth distribution of the oxygen ligands measured by angle-resolved XPS, a bilayer model was built for quantitative processing of the XPS measured intensities [27]. The model, shown in Fig. 4, assumes a mixed chromium-iron oxide inner layer, neglecting any molybdenum content, and a mixed iron-chromium-molybdenum oxide/hydroxide outer layer. The equivalent thickness and composition of the outer and inner layers and the composition of the alloy underneath were calculated from the intensities of the oxidized and metallic components used for reconstruction: Cr2-Cr6 assigned to inner layer and Cr7 assigned to outer layer, Fe2-Fe4 assigned to inner layer and Fe5-Fe7 assigned to outer layer, Mo2-Mo3 assigned to outer layer, and Cr1, Fe1, Mo1 and Ni1 all assigned to modified alloy substrate. A classical model of intensity attenuation by continuous layers was used. The results are presented in Table. 2. The total thickness is found to be 2 nm, slightly larger than that (1.7 nm calculated using a similar model) reported for the native oxide film formed in air on a Fe–17Cr–14.5Ni–2.3Mo (wt%) single-crystalline surface prepared by mechanical polishing, electropolishing and high temperature annealing in reducing hydrogen atmosphere [27]. This difference might be due, at least partially, to the different surface preparation and the remaining presence of a cold-worked layer on the polycristalline mechanically polished surface. In the present case , the thicknesses of the outer and inner layers of the surface oxide are 0.8 and 1.2 nm, respectively. Their ratio of 2/3 is in agreement with the value of ~8/12 obtained from sputtering time in the ToF-SIMS depth profiles, not far from the 41/59 value found on the single-crystalline surface [27].

The overall composition of the oxide film, obtained by weighting the cation concentration value of each element by the fractional thickness of the inner and outer layers, as well as the composition of the modified alloy beneath the oxide film are also shown in Table. 2. Chromium is found markedly enriched with $Cr^{III}$ ions representing 55% of the metal cations in the film and slightly depleted (18 at% instead of 20 at% in the bulk alloy) in the modified alloy underneath the oxide film. The overall Cr/Fe ratio in the oxide film is 1.4 vs. 0.3 in the modified alloy. This ratio is found higher in the inner layer than in the outer layer of the oxide film, reproducing the trend evidenced by the ToF-SIMS depth profiles. Nickel, not found in measurable amount in the oxide film, is enriched in the alloy underneath the oxide film (28 at% instead of 12 at% in the bulk alloy), also in agreement with the ToF-SIMS depth profiles. These enrichments are quite similar to the results for the single-crystalline surface [27]. Concerning molybdenum, it is found enriched in the native oxide film (6 at% of $Mo^{IV/VI}$ instead of 1.6 at% of $Mo^0$ in the bulk alloy), more significantly than on the single-crystalline surface (2 at%) but also enriched in the alloy underneath the oxide film (3 at%), in contrast to the absence of enrichment found on the single-crystalline surface (1.5 at%), possibly also owing to the different surface preparation and/or microstructure effects.

### 3.1.2 Passive films

The ToF-SIMS negative ion depth profiles for the 316L stainless steel samples passivated in 0.05 M $H_2SO_4$(aq) are shown in Fig. 5 (1 h) and Fig. 6 (20 h). In the surface oxide regions, the most predominant profiles of the oxidized metallic elements remain those of the $CrO_2^-$, $FeO_2^-$ and to a lesser extent $MoO_2^-$ ions for both passive films (Fig. 5a and Fig. 6a). Like for the native oxide film, the $NiO_2^-$ ions have much lower intensity. Using the $Ni_2^-$ ions profiles and the same criterion as for the native oxide-covered sample, the metallic substrate regions are positioned at 49 s for both passive films. The oxide film is from 0 to 19 s for the 1 h passive film and from 0 to 20 s for the 20 h passive film, suggesting no marked variation of the overall thickness compared to the native oxide film. For both passive films, the modified alloy region is enriched in Ni as shown by the peaks (attenuated in logarithmic scale) observed in the $Ni_2^-$ ions profiles at the beginning of this region. This enrichment is confirmed by XPS (Table 1).

For the 1 h passivated sample, the profiles of the $FeO_2^-$ and $CrO_2^-$ secondary ions and $FeO_2^-$ and $MoO_2^-$ secondary ions are compared in Fig. 5b and Fig. 5c, respectively, using the same intensity scales as in Fig. 1. The $FeO_2^-$ and $CrO_2^-$ profiles also peak at different positions, 5 and 10 s, respectively, showing a bilayer structure also for the passive film with Fe and Cr oxides more concentrated in the outer and inner layers, respectively. By using the same method, we assign the outer/inner layer interface at 7 s. The $FeO_2^-$ and $MoO_2^-$ profiles both peak at 5 s and superimpose at longer sputtering time (Fig. 5c). The partition of the passive film is confirmed in Fig. 5d presenting the variation of the intensity ratio of $CrO_2^-$ to $FeO_2^-$ ions. The tendency of the curve is the same as for the native oxide film but with a higher Cr/Fe ratio, meaning that the passive oxide film has a higher chromium enrichment both in the outer and inner layers.

For the 20 h passivated sample, the $FeO_2^-$ and $CrO_2^-$ profiles, shown in Fig. 6b, also peak at different positions, 5 and 8 s, confirming the bilayer structure for the 20 h passive film with Fe and Cr oxides more concentrated in the outer and inner layers, respectively (interface is placed at 6 s). The $FeO_2^-$ and $MoO_2^-$ profiles both peak at 5 s and superimpose at longer sputtering time (Fig. 6c). The partition of the passive film is confirmed in Fig. 6d presenting the variation of the intensity ratio of $CrO_2^-$ to $FeO_2^-$ ions. The Cr/Fe ratio is slightly higher than that for the 1 h passive film, suggesting further Cr enrichment under the effect of aging in the passive state.

Fig. 7 compares the ToF-SIMS data for native oxide-covered and passivated samples. As expected under identical analytical conditions of the same substrate in all samples, the intensity of the $Fe_2^-$ ions is the same in the metallic substrate region beyond 50 s (Fig. 7a). This means that no normalization is necessary to compare the profiles of the three samples. The shift towards shorter sputtering time of the increase in intensity of the $Fe_2^-$ ions reflects the decrease in thickness of the surface oxide film; it is noticeable for the 20 h passive film only. The Cr content in the inner layer is higher on the two passivated samples, indicating increased Cr oxide enrichment after electrochemical passivation and a promoting effect of aging (Fig. 7b). In the outer layer, the Fe content is slightly decreased (Fig. 7c), indicating that preferential iron oxide dissolution in acid solution is the cause of the increasing Cr enrichment. A small increase of Mo oxide content in the outer layer is suggested after 20 h passivation but not confirmed after 1 h (Fig. 7d).

The XPS spectra recorded for the 1 h passivated sample are similar to those for the native oxide-covered sample (Fig. 8). The fitting results are also presented in Table. 1. The Cr $2p_{3/2}$ components associated to metallic $Cr^0$ in the substrate (Cr1), $Cr^{III}$ oxide (Cr2-Cr6) and $Cr^{III}$ hydroxide (Cr7) in the passive film are observed at 574.1 eV, 576.2-579.4 eV and 577.2 eV respectively (Fig. 8a) [13,16,33,48,49]. The Fe $2p_{3/2}$ components associated to metallic $Fe^0$ in the substrate (Fe1), $Fe^{III}$ oxide (Fe2-Fe6) and $Fe^{III}$ hydroxide (Fe7) in the passive film are observed at 707 eV, 708.8-712.8 eV and 711.9 eV respectively (Fig. 8b) [7,22,23,34,48,51,52]. The Mo $3d_{5/2-3/2}$ doublet components associated to metallic $Mo^0$ in the substrate and $Mo^{IV}$ and $Mo^{VI}$ in the passive film are observed at 227.8-230.9 eV, 229.6-232.7 eV and 232.7-235.8 eV, respectively (Fig. 8c) [31,33,36,53]. Since passivation was performed in sulfuric acid, a S 2s peak at 232.6 eV had to be considered. The presence of sulfate was confirmed by the S 2p spectrum (signal at 168.9 eV not shown here). The intensity ratio of the $Mo^{VI}$ to $Mo^{IV}$ doublets is ~6. The three O 1s components assigned to the oxide ($O^{2-}$), hydroxide ($OH^-$) and water ($H_2O$) ligands in the oxide film are observed at 530.3, 531.8 and 532.7 eV, respectively (Fig. 8d) [11,34]. The intensity ratio of $OH^-/O^{2-}$ is 1.2 for the 1 h passivated sample, higher than for the native oxide-covered sample (0.9). It was confirmed by comparing the C 1s spectra that the increase of this component is not related to a higher carbonate contamination on the passivated sample, so that it can be concluded that the passive film is indeed more hydroxylated than the native oxide film as expected after formation in the aqueous solution.

Fig. 9 shows the XPS spectra for the 20 h passivated sample. The fitting results are presented in Table. 1. The BE positions and FWHM values of component peaks are identical within ±0.1 eV to those for the 1 h passive and native oxide films. Only slight changes of the relative intensities are observed. Noteworthy is the slight decrease of the $OH^-/O^{2-}$ intensity ratio to 1.1, indicating that dehydroxylation of the passive film is promoted by aging in the passive state in agreement with previous studies [2,20,27]

Using the same model of the bilayer structure (Fig. 4), thickness and composition of the 1 h and 20 h passive films were calculated (Table. 2). Compared with the native oxide film, the slight decrease of the thickness suggested by ToF-SIMS is confirmed and concentrated in the outer layer. We calculate values of 0.7 nm for 1 h and 0.6 nm for 20 h passivation instead of 0.8 nm for the native oxide. The composition of outer layer is calculated as 26%$Fe^{III}$-56%$Cr^{III}$-18%$Mo^{IV/VI}$ for 1 h and 32%$Fe^{III}$-49%$Cr^{III}$-19%$Mo^{IV/VI}$ for 20 h passivation instead of 41%$Fe^{III}$-44%$Cr^{III}$-15%$Mo^{IV/VI}$ for the native oxide, confirming chromium and molybdenum enrichment after passivation. The inner layer composition is 26%$Fe^{III}$-74%$Cr^{III}$ for 1 h and 23%$Fe^{III}$-77%$Cr^{III}$ for 20 h passivation instead of 36% $Fe^{III}$-64% $Cr^{III}$ for the native oxide. It reproduces the increased balance in favor of chromium also observed by ToF-SIMS (Fig. 5d and Fig. 6d) and reflected by the increase of the Cr/Fe ratio calculated from the XPS data (Table. 2). The global composition of the film also reflects the increase of the chromium enrichment induced by passivation. The $Cr^{III}$ concentration increases to 67% in the oxide film after 1 h passivation versus 55% in the native oxide but remains constant after 20 h passivation (68%). The passivation-induced $Mo^{IV/VI}$ enrichment of the oxide film outer layer is not marked when considering the global composition of the oxide. The alloy underneath the oxide film is found still markedly enriched in nickel and only slightly in molybdenum after passivation. The slight chromium depletion observed underneath the native oxide seems vanished after 1 h passivation and would reappear after longer passivation.

These modifications of the oxide film composition are consistent with the preferential iron oxide dissolution induced by passivation in acid solution [16,27,30,32,37]. Comparing with the results reported for the single-crystalline surface [27], the differences observed in the present case are the absence of variation of thickness of the inner layer of the oxide film after passivation whereas an increase was observed on the single-crystalline surface. The decrease in chromium depletion in the alloy underneath the film observed after 1 h passivation was also observed on the single-crystalline surface. The effect of aging in the passive state is to increase the Cr enrichment in the inner layer of the oxide film, considered as the barrier layer, but not in the outer layer, considered as an exchange layer with the electrolyte. It confirms previous findings on Mo-free austenitic [20] and ferritic [2,12] SS. Concerning molybdenum, its enrichment in the outer layer as well as its overall concentration in the oxide film do not seem to be promoted by aging in the passive state, nor its enrichment in the modified alloy underneath the passive film. The confinement of Mo$^{IV/VI}$ enrichment in the outer layer of the passive film is in agreement with previous works in the presence of molybdates [22,30,32]. It is supportive of the bipolar model developed to explain how molybdenum would mitigate passive film breakdown [34-36].

## 3.2. Photoelectrochemical characterization

### 3.2.1 Native oxide film

The photoelectrochemical study was conducted in 0.1 M $Na_2B_4O_7$ electrolyte (pH ~ 9.5) at open circuit potential $U_{OC}$ = -0.07 V vs Ag/AgCl and at slightly higher potential ($U_E$ = 0.2 V vs Ag/AgCl in order to minimize the changes in the oxide film composition induced by anodic polarization. Fig. 10a shows the photocurrent spectrum recorded at $U_{OC}$ on the native oxide film. The decrease of the photocurrent efficiency to noise level with increasing wavelength (decreasing energy) is consistent with the presence of a surface oxide film with an optical band gap at the interface between electrolyte and metal substrate. It is possible to estimate a band gap value using Eq. (1), valid for an indirect optical transition and photon energies close to the band gap value [54]:

$$(Q_{ph} h\nu)^n \propto (h\nu - E_g) \qquad (1)$$

where h$\nu$ is the photon energy and $E_g$ is the optical band gap (or mobility gap for amorphous materials); n is assumed to be 0.5 for non-direct optical transitions [54,55]. The photocurrent yield $Q_{ph}$ is assumed to be proportional to the light absorption coefficient. It is defined as:

$$Q_{ph} = \frac{I_{ph}}{e\phi_0(1-R)} \qquad (2)$$

where $I_{ph}$ is the collected photocurrent and e the electron charge. The incident photon flux on the interface $\Phi_0$ is corrected for the reflection at the metal/oxide/electrolyte interface with R being the total reflectivity of the junction.

The inset of Fig. 10a shows the $(Q_{ph} \cdot h\nu)^{0.5}$ vs $h\nu$ plot. Two linear regions are evidenced. From the line fitting of the high energy region, one estimates an optical band gap close to 3.0 eV by extrapolating the $(Q_{ph}h\nu)^{0.5}$ value to the noise level set at $1\times10^{-3}$ a.u.. Using the same procedure, a band gap close to 2.7 eV is estimated from the line fitting of the low energy region.

Fig. 10b shows the photocurrent spectrum recorded at $U_E = 0.2$ V vs Ag/AgCl for the same native oxide sample. No marked difference is observed between the two spectra. Also in this case, the $(Q_{ph} \cdot h\nu)^{0.5}$ vs $h\nu$ plot evidences two linear regions (see inset of Fig. 10b), with band gap values estimated to 3.0 eV and 2.6 eV from the high and low energy regions, respectively, in agreement with the values determined at open circuit potential. Thus, anodic polarization at 0.2 V vs Ag/AgCl does not induce marked alterations of the passive film, as suggested by the absence of variation of the band gap values. The partition in two linear regions of the $(Q_{ph} \cdot h\nu)^{0.5}$ vs $h\nu$ plots can be associated with the bilayer structure of the oxide film evidenced by ToF-SIMS and confirmed by XPS, owing to the different chemical composition of the outer and inner layers. A two-region partition of the photocurrent spectra was also found for thermal oxide films formed on bright-annealed ferritic stainless steel [46], and on passive films formed on austenitic stainless after immersion in high pressurized water [45] and after electrochemically-induced rouging [47]. A bilayer structure of the surface oxide films with outer and inner layers of different chemical composition and associated to different band gap values was also formed in these cases.

The plots of the photocurrent efficiency and phase recorded versus applied potential at a wavelength of 309 nm are shown in Fig. 10c. The marked drop of the photocurrent at $U_E \geq 0.2$ V vs Ag/AgCl can be the result of the very high electric field modifying the surface oxide film in this potential range. At lower applied potentials, the photocurrent variation is inversed when shifting in the cathodic direction. The photocurrent first decreases between 0.2 and -0.4 V, as typical for anodic photocurrent, and then increases between -0.4 and -0.8 V as typical for cathodic photocurrent. The variation of the photocurrent phase shows a transition centered at about -0.4 V, from about 10° in the anodic region to about -150° in the cathodic region. This behavior characterizes an insulating material, for which depending on the applied potential with respect to the flat band potential and thus on the direction of the electric field both anodic and cathodic photocurrents can be measured [47]. Since for insulating layers the flat band potential can be assumed to be close to the inversion photocurrent potential, we can estimate the flat band potential of the native oxide film formed on 316L SS to be about - 0.4 V vs Ag/AgCl at pH ~9.5. This value is in agreement with the value reported for passive films grown on pure chromium [55]. Due to the very low thickness of the oxide film, we cannot exclude that the cathodic photocurrent arises from electron photoemission processes [45]. In this case the oxide would behave as a thin n-type semiconductor.

In order to confirm the sign of the generated photocurrent, current vs time plots were recorded at different $U_E$ values under constant wavelength while manually chopping the irradiation (Fig. 11). The generated photocurrent changes from anodic to none when the applied potential changes from 0.1 to -0.4 V vs Ag/AgCl. The drop to zero at $U_E$= -0.4 V is consistent with the data reported in Fig. 7c and the determination of the flat band potential. The cathodic current expected at lower applied potential cannot be detected because the background current is too high.

### 3.2.2 Passive films

The photocurrent spectra for the austenitic stainless steel samples passivated in 0.05 M $H_2SO_4$ at $U_{Pass}$= 0.3 V/SCE are presented in Fig. 12. They were obtained in 0.1 M $Na_2B_4O_7$ at $U_E$ = 0.2 V vs Ag/AgCl and can thus be directly compared to that for the native oxide film. For both passivated samples, the photocurrent efficiency is lower than for the native oxide-covered sample mostly at low wavelength (high energy), meaning that the passive films are less photoactive in this region associated with the inner layer of the oxide films. The low measured photocurrents prevent a reliable determination of the band gap as made for the native oxide film.

According to the XPS and ToF-SIMS results, the main passivation-induced alterations of the oxide film are (i) slight decrease in thickness, (ii) higher hydroxylation of the oxide matrix, (iii) Mo enrichment and (iv) Cr enrichment. The slight decrease in thickness affects the outer layer as a result of preferential iron oxide dissolution induced by passivation. Hydroxylation is also mostly concentrated in the outer layer as shown by the angle-resolved XPS data (Fig. 2e). As for the Mo enrichment, it increases in the outer layer as discussed above. In contrast, the Cr enrichment increases in both layers after passivation as discussed and it may be at the origin of the loss of photo-activity observed after passivation. Besides, its slight increase with passivation time is also in agreement with increase loss of photo-activity after 20 h passivation.

### 3.2.3 Modeling

In order to get a more precise estimate of the optical band gap and photo-generated carriers transport properties of the two in-series layers of the oxide films, we have applied a model previously proposed to simulate the photoelectrochemical behavior of bilayered films [56,57]. According to this model, the collected photocurrent can be calculated as the sum of the contributions coming from each layer for the bilayered structure. The following equation is used for interpolating the experimental data:

$$(Q \cdot h\nu)^{1/2} = \left[\left(T_{out}G_{out} + e^{-\alpha_{out}D_{out}} \times T_{inn}G_{inn}\right)h\nu\right]^{1/2} \qquad (3)$$

In this equation, the subscripts "inn" and "out" refer to the inner and outer layers, respectively, G and T to the generation and transport term in each layer, respectively, $\alpha_{out}$ to the absorption coefficient of photon by the outer layer, and $D_{out}$ to the thickness of the outer layer. The photon flux impinging the inner layer surface is assumed to be $\Phi_0 e^{-\alpha_{out}D_{out}}$.

The expressions for the G and T terms are [57]:

$$G = \eta_g \left[1 - \exp(-\alpha D)\right] \qquad (4)$$

and

$$T = \frac{\mu\tau F}{D}\left[1 - \exp\left(-\frac{D}{\mu\tau F}\right)\right] \qquad (5)$$

where $\eta_g$ ($\lambda$, F) is the generation efficiency, lower than 1 as result of geminate recombination effects in disordered materials at low electrode potentials with respect to the flat band potential [58]. $\alpha$ is the absorption coefficient at the given wavelength, D the total film thickness (D = $D_{out}$ + $D_{inn}$) and F the electric field inside the film. $\mu$ and $\tau$ are the mobility and lifetime of the photogenerated carriers (holes, h, and electrons, e), respecting the following equation:

$$\mu\tau = \mu_e\tau_e + \mu_h\tau_h \qquad (6)$$

In order to take into account possible multiple reflection effects at the metal/oxide interface, an average value of the total reflectivity, R, of the electrolyte/surface oxide film/metal junction has been estimated under the hypothesis of a single absorbing layer on an absorbing substrate [59]. The refractive indexes relating to water [60], the composition of surface oxide [61] and stainless steel [62] were used.

Fig. 13 presents the experimental $(Q \cdot h\nu)^{0.5}$ vs $h\nu$ plots and their fit according to Eq. (4) for the native oxide film. The $D_{out}$ and $D_{inn}$ values obtained by XPS have been used. The fitted values of α, in the range $10^5$ - $10^3$ cm$^{-1}$ for the Fe-rich outer layer and $10^4$ - $10^2$ cm$^{-1}$ for the Cr-rich inner layer are in quite good agreement with the values usually reported for iron and chromium oxides [63-64]. Table 3 presents the other fitting parameters. For the native oxide film, the band gap of the inner layer is 3.3 eV and the band gap of the outer layer is 2.4 eV. These two values are in reasonable good agreement with those obtained by linear fitting of the $(Q_{ph} \cdot h\nu)^{0.5}$ vs $h\nu$ plots (Fig. 10). For the passive films, no fits of the experimental curves could be achieved due to the weak measured photo currents (Fig. 12).

According to previous works [64-66], the bandgap value of mixed oxides $A_aB_bO_o$ depends on the electronegativity of the oxide. For d-metal oxides, the relationship is the following:

$$E_g - \Delta E_{am} \text{ (eV)} = 1.35 (\chi_{av} - \chi_O)^2 - 1.49 \tag{7}$$

where $\chi_O$ is the oxygen electronegativity (3.5 in the Pauling scale) and $\chi_{av}$ an average electronegativity parameter defined as the arithmetic mean between the electronegativity of the metal partners in the oxides, i.e.:

$$\chi_{av} = \frac{a}{a+b}\chi_A + \frac{b}{a+b}\chi_B \tag{8}$$

$\Delta E_{am} = 0$ for crystalline oxides, whilst increasing values are expected (up to around 0.5 eV) if the lattice disorder affects the density of states distribution both near the valence and conduction band edges [58].

In our case, we can develop Eq. (9) based on Eq. (8):

$$\chi_{av} = \frac{a}{a+b+c}\chi_A + \frac{b}{a+b+c}\chi_B + \frac{c}{a+b+c}\chi_c \tag{9}$$

For the native oxide film the composition of the outer and inner layers are taken as 41%Fe-44%Cr-15%Mo and 36%Fe-64%Cr, respectively, from the XPS data (Table 2). We assume $\chi\text{Fe}^{3+} = 1.9 \pm 0.05$, $\chi\text{Cr}^{3+} = 1.6 \pm 0.05$ and $\chi\text{Mo}^{6+} = 1.65 \pm 0.05$. Using Eq. (7) and Eq. (9) we can calculate the theoretical values of the band gaps and, considering the electronegativity uncertainty, obtain values of $2.7 \pm 0.3$ eV and $2.8 \pm 0.3$ eV for the outer and inner layers, respectively (considering $\Delta E_{am}$ is 0 eV). These values are in reasonable good agreement with the values obtained by the modeling of the experimental data of Fig. 10. For the passive films, the theoretical values of the band gaps can also be calculated from the composition of the layers determined by surface analysis. Considering the compositions of 26%Fe-56%Cr-18%Mo and 23%Fe-77%Cr for the outer and inner layers of the 1 h passive film, we obtain values of 2.9 ± 0.3 eV and 3.0 ± 0.3 eV, respectively, reflecting the effect of the increased Cr enrichment on the expected band gap values. For the 20 h passive film, the expected respective values would also be 2.9 ± 0.3 eV and 3.0 ± 0.3 eV.

### 3.3. Electrochemical characterization

Fig 14a compares the polarization curves recorded in 0.1 M $Na_2B_4O_7$ electrolyte (pH ~9.5) at 1 mV s$^{-1}$ for the native oxide-covered and 1 h pre-passivated samples. On both samples, no active-passive transition is observed in agreement with the presence of Cr-enriched surface oxides films. The passivity region is stable up to 0.4 V vs Ag/AgCl, suggesting that transpassivity associated with Cr(VI) oxide formation and dissolution is not reached in these conditions. The curves confirm that anodic polarization in 0.1 M $Na_2B_4O_7$ does not modify markedly the passive film composition, as supported by the photoelectrochemical characterizations performed at 0.2 V. The PCS data recorded at 0.4 V (Fig 10c) suggest however some modifications of the native oxide film. The corrosion potential is shifted toward the anodic direction for the pre-passivated sample and the passive current density is lower, meaning that the passive oxide film produced by polarization in acid solution offers better corrosion protection than the native oxide film in agreement with the enhanced Cr-enrichment measured by surface analysis.

The potential $U_E$ = 0 V vs Ag/AgCl was selected to record electrochemical impedance spectra based on the polarization curves. Fig 14b presents the EIS data in Nyquist representation. A portion of a deformed semicircle describes the dependence of the imaginary vs real components of the impedance [67]. This has been discussed as related to the influence of the frequency on the overall surface oxide capacitance caused by gap states for amorphous or strongly disordered thin oxide layers [58] and by surface states and adsorption phenomena [47]. The simplified equivalent circuit shown in the inset was used to fit the EIS data. It introduces a constant phase element to model the frequency-dependent equivalent capacitance and to calculate the polarization resistance $R_P$ as a rough estimate of the corrosion resistance. The values of best fit parameters are compiled in Table. 4. The polarization resistance increases for the pre-passivated samples whereas the other parameters remain essentially unchanged. This means that despite the passivation-induced slight thickness decrease of the surface oxide film, evidenced by surface analysis, the corrosion resistance is higher than for the native oxide-covered sample film in agreement electrochemical polarization measurements. This confirms the effect associated with the enhanced Cr-enrichment of the surface oxide measured by surface analysis. No effect of aging in the passive state is detected on the polarization resistance.

## 4. Conclusion

ToF-SIMS, XPS, PCS and electrochemical measurements were combined in order to study the physicochemical alterations induced by electrochemical passivation of the surface oxide film providing corrosion resistance to 316L austenitic stainless steel. The native oxide film formed on the mechanically polished SS surface is ~2 nm thick and consists of a mixed Cr(III)-Fe(III) oxide hydroxylated in its outer part. It has a bilayer structure highly enriched in Cr(III) in the inner layer and less so in the outer layer. Molybdenum is concentrated, mostly as Mo(VI), in the outer layer and nickel is below the XPS detection limit. Photocurrent spectroscopy performed *in situ* in a borate buffer solution at and near open circuit potential confirmed the bilayer structure. The band gaps of the inner and outer layers are 3.0 eV and 2.6-2.7 eV, respectively, in agreement with the variation of the chromium enrichment. According to the PCS data, the oxide film would behave as an insulator with an inversion potential of ~ - 0.4 V vs. Ag/AgCl at pH ~ 9.5.

Electrochemical passivation in sulfuric acid solution causes the Cr(III) enrichment to increase in both layers of the oxide film and the Mo(IV-VI) enrichment to increase in the outer layer, owing to the preferential dissolution of Fe(III) in the electrolyte also reflected by the slight decrease of thickness of the outer layer of the oxide film. The decrease of the photoactivity measured *in situ* prevented a reliable experimental determination of the band gaps of the inner and outer layers after passivation but values of 3.0 and 2.9 eV, respectively, were calculated from the composition of the layers measured by XPS. Electrochemical passivation and the related enrichment in chromium of the oxide film improves the corrosion resistance as evidenced by the anodic shift of the corrosion potential and the increase of the polarization resistance observed on the pre-passivated surface by linear sweep voltammetry and impedance spectroscopy. Aging in the passive state promotes the Cr enrichment in the inner barrier layer of the passive film, however with no detectable effects on the polarization resistance. It also promotes dehydroxylation, thereby counter-acting the hydroxylation increase induced by passivation in aqueous electrolyte.

## Acknowledgements

This project has received funding from the European Research Council (ERC) under the European Union's Horizon 2020 research and innovation program (ERC Advanced Grant no. 741123). Région Île-de-France is acknowledged for partial funding of the ToF-SIMS equipment.

# Figure captions

Fig. 1 ToF-SIMS depth profiles for the native oxide film on 316L SS: (a) $^{18}O^-$, $^{18}OH^-$, $CrO_2^-$, $FeO_2^-$, $NiO_2^-$, $MoO_2^-$, $Cr_2^-$, $Fe_2^-$ and $Ni_2^-$ secondary ions, (b) $CrO_2^-$ and $FeO_2^-$ secondary ions, (c) $FeO_2^-$ and $MoO_2^-$ secondary ions, (d) $CrO_2^-/FeO_2^-$ intensity ratio.

Fig.2 Angle-resolved XPS core level spectra recorded at 40° (red line) and 90° (black line) take-off angle for the native oxide film on 316L SS: (a) Cr $2p_{3/2}$, (b) Fe $2p_{3/2}$, (c) Mo $3d_{5/2-3/2}$, (d) Ni $2p_{3/2}$ and (e) O 1s regions. The spectra were normalized in intensity at background level.

Fig. 3 XPS core level spectra and their reconstruction for the native oxide film on 316L SS: (a) Cr $2p_{3/2}$, (b) Fe $2p_{3/2}$, (c) Mo $3d_{5/2-3/2}$ and (d) O 1s regions (take-off angle: 90°).

Fig. 4 Model of duplex oxide film on modified alloy substrate used for calculating thickness and composition from the XPS data

Fig. 5 ToF-SIMS depth profiles for 316L stainless steel passivated in 0.05 M $H_2SO_4$ at $U_E$= 0.3 V/SCE for 1 h : (a) $^{18}O^-$, $^{18}OH^-$, $CrO_2^-$, $FeO_2^-$, $NiO_2^-$, $MoO_2^-$, $Cr_2^-$, $Fe_2^-$ and $Ni_2^-$ secondary ions, (b) $CrO_2^-$ and $FeO_2^-$ secondary ions, (c) $FeO_2^-$ and $MoO_2^-$ secondary ions, (d) $CrO_2^-/FeO_2^-$ intensity ratio.

Fig. 6 ToF-SIMS depth profiles for 316L stainless steel passivated in 0.05 M $H_2SO_4$ at $U_E$= 0.3 V/SCE for 20 h : (a) $^{18}O^-$, $^{18}OH^-$, $CrO_2^-$, $FeO_2^-$, $NiO_2^-$, $MoO_2^-$, $Cr_2^-$, $Fe_2^-$ and $Ni_2^-$ secondary ions, (b) $CrO_2^-$ and $FeO_2^-$ secondary ions, (c) $FeO_2^-$ and $MoO_2^-$ secondary ions, (d) $CrO_2^-/FeO_2^-$ intensity ratio.

Fig.7 Comparison of ToF-SIMS depth profiles for the native and passive oxide films: (a) $Fe_2^-$ secondary ions, (b) $CrO_2^-$ secondary ions, (c) $FeO_2^-$ secondary ions, (d) $MoO_2^-$ secondary ions

Fig. 8 XPS core level spectra and their reconstruction for 316L SS passivated in 0.05 M $H_2SO_4$ at $U_E$= 0.3 V/SCE for 1 h: (a) Cr $2p_{3/2}$, (b) Fe $2p_{3/2}$, (c) Mo $3d_{5/2-3/2}$ and (d) O 1s regions (take-off angle: 90°).

Fig. 9 XPS core level spectra and their reconstruction for 316L SS passivated in 0.05 M $H_2SO_4$ at $U_E$= 0.3 V/SCE for 20 h: (a) Cr $2p_{3/2}$, (b) Fe $2p_{3/2}$, (c) Mo $3d_{5/2-3/2}$ and (d) O 1s regions (take-off angle: 90°).

Fig. 10 PCS analysis in 0.1 M $Na_2B_4O_7$ of the native oxide film on 316L SS: (a,b) photocurrent spectra and $(Q_{ph}h\nu)^{0.5}$ vs $h\nu$ plots (insets) at (a) $U_{OCP}$ = - 0.07 V vs Ag/AgCl and (b) at $U_E$= 0.2 V vs Ag/AgCl; (c) photocurrent and phase vs applied potential for $\lambda$ = 309 nm.

Fig. 11 Photocurrent vs time as measured in 0.1 M $Na_2B_4O_7$ under polarization without (D) and with (L) illumination (λ=309 nm) on the native oxide-covered 316L SS sample: (a) $U_E$= 0.1 V vs Ag/AgCl. b) $U_E$= -0.2 V vs Ag/AgCl c) $U_E$= -0.4 V vs Ag/AgCl.

Fig. 12 PCS analysis in 0.1 M $Na_2B_4O_7$ for 316L stainless steel passivated in 0.05 M $H_2SO_4$ at $U_E$= 0.3 V/SCE for 1 h and 20 h and comparison with the native oxide-covered sample.

Fig. 13 Fit (continuous line), according to eq. (3), of the experimental $(Qh\nu)^{1/2}$ vs h$\nu$ plots (points) for the native oxide film

Fig. 14 Comparative electrochemical analysis of native oxide-covered and pre-passivated 316L SS: (a) polarization curves recorded in 0.1 M $Na_2B_4O_7$ (pH 9.5) at 1 mV s$^{-1}$ scan rate; (b) EIS spectra (Nyquist plots) recorded in 0.1 M $Na_2B_4O_7$ (pH 9.5) at 0 V vs Ag/AgCl. The inset shows the equivalent circuit used for fitting the data.

# Tables

Table. 1 BE, FWHM and relative intensity values of the peak components used for reconstruction of the XPS spectra on polycrystalline 316L SS after formation in ambient air of the native oxide film and after passivation at 0.3 V/SCE in 0.05 M $H_2SO_4$.

| Core level | Peak | Assignment | Native oxide film | | | Passive oxide film 1 h | | | Passive oxide film 20 h | | |
|---|---|---|---|---|---|---|---|---|---|---|---|
| | | | BE (±0.1 eV) | FWHM (±0.1 eV) | Intensity (%) | BE (±0.1 eV) | FWHM (±0.1 eV) | Intensity (%) | BE (±0.1 eV) | FWHM (±0.1 eV) | Intensity (%) |
| Fe $2p_{3/2}$ | Fe1 | Fe metallic | 707.0 | 0.8 | 55.1 | 707.0 | 0.8 | 63.3 | 707.0 | 0.8 | 63.8 |
| | Fe2 | $Fe^{III}$ oxide | 708.8 | 1.3 | 8.9 | 708.8 | 1.2 | 8.2 | 708.9 | 1.2 | 8.5 |
| | Fe3 | $Fe^{III}$ oxide | 709.9 | 1.1 | 8.0 | 709.8 | 1.1 | 7.4 | 709.9 | 1.1 | 7.6 |
| | Fe4 | $Fe^{III}$ oxide | 710.7 | 1.1 | 6.3 | 710.7 | 1.1 | 5.7 | 710.8 | 1.1 | 5.9 |
| | Fe5 | $Fe^{III}$ oxide | 711.7 | 1.3 | 3.6 | 711.7 | 1.3 | 3.3 | 711.8 | 1.3 | 3.4 |
| | Fe6 | $Fe^{III}$ oxide | 712.8 | 2.0 | 3.6 | 712.8 | 2.0 | 3.3 | 712.9 | 2 | 3.4 |
| | Fe7 | $Fe^{III}$ hydroxide | 711.9 | 2.7 | 14.5 | 711.9 | 2.7 | 8.8 | 711.9 | 2.7 | 7.5 |
| Cr $2p_{3/2}$ | Cr1 | Cr metallic | 574.1 | 1.1 | 26.4 | 574.1 | 1.1 | 24.5 | 574.0 | 1.1 | 21.2 |
| | Cr2 | $Cr^{III}$ oxide | 576.0 | 1.2 | 17.8 | 576.2 | 1.4 | 17.4 | 576.1 | 1.3 | 19.0 |
| | Cr3 | $Cr^{III}$ oxide | 577.0 | 1.2 | 17.2 | 577.2 | 1.4 | 16.9 | 577.1 | 1.3 | 18.4 |
| | Cr4 | $Cr^{III}$ oxide | 577.8 | 1.2 | 9.4 | 578..0 | 1.4 | 9.2 | 577.9 | 1.3 | 10.1 |
| | Cr5 | $Cr^{III}$ oxide | 578.8 | 1.2 | 39 | 579.0 | 1.4 | 3.8 | 578.9 | 1.3 | 4.2 |
| | Cr6 | $Cr^{III}$ oxide | 579.2 | 1.2 | 2.5 | 579.4 | 1.4 | 2.4 | 579.3 | 1.3 | 2.6 |
| | Cr7 | $Cr^{III}$ hydroxide | 577.2 | 2.5 | 22.8 | 577.2 | 2.5 | 25.8 | 577.2 | 2.5 | 24.4 |
| Mo $3d_{5/2}$ | Mo1 | Mo metallic | 227.7 | 0.5 | 28.4 | 227.8 | 0.5 | 28.6 | 227.7 | 0.5 | 27.4 |
| | Mo2 | $Mo^{IV}$ oxide | 229.5 | 0.8 | 3.4 | 229.6 | 0.9 | 4.6 | 229.5 | 0.9 | 3.6 |
| | Mo3 | $Mo^{VI}$ oxide | 232.6 | 2.3 | 28.4 | 232.7 | 2.5 | 27.1 | 232.6 | 2.5 | 29.3 |
| Mo $3d_{3/2}$ | Mo1' | Mo metallic | 230.9 | 0.8 | 18.8 | 230.9 | 0.9 | 18.8 | 230.8 | 0.8 | 18.1 |
| | Mo2' | $Mo^{IV}$ oxide | 232.6 | 0.8 | 2.3 | 232.7 | 0.9 | 3.1 | 232.7 | 0.9 | 2.4 |
| | Mo3' | $Mo^{VI}$ oxide | 235.7 | 2.3 | 18.8 | 235.8 | 2.5 | 17.9 | 235.6 | 2.5 | 19.3 |
| Ni $2p_{3/2}$ | Ni | Ni metallic | 852.9 | 0.9 | 100 | 852.9 | 0.9 | 100 | 852.8 | 0.9 | 100 |
| O 1s | O1 | $O^{2-}$ | 530.3 | 1.2 | 43.1 | 530.3 | 1.2 | 33.7 | 530.3 | 1.2 | 40.5 |
| | O2 | $OH^-$ | 531.6 | 1.7 | 39.9 | 531.8 | 1.7 | 39.5 | 531.7 | 1.8 | 44.2 |
| | O3 | $H_2O$ | 532.8 | 2.4 | 17 | 532.7 | 2.5 | 26.9 | 532.8 | 2.5 | 15.3 |

Table. 2 Thickness and composition of the native and passive oxide films on polycrystalline 316L SS as calculated form the XPS data based on the model in Fig. 4.

| | | Global film | Outer layer | Inner layer | Modified alloy |
|---|---|---|---|---|---|
| **Native film** | d (nm) | 2 | 0.8 | 1.2 | / |
| | [Fe] (at%) | 39 | 41 | 36 | 52 |
| | [Cr] (at%) | 55 | 44 | 64 | 18 |
| | [Ni] (at%) | / | / | / | 28 |
| | [Mo](at%) | 6 | 15 | / | 3 |
| | Ratio Cr/Fe | 1.4 | 1.1 | 1.8 | 0.3 |
| **Passive film 1h** | d (nm) | 1.9 | 0.7 | 1.2 | / |
| | [Fe] (at%) | 26 | 26 | 26 | 51 |
| | [Cr] (at%) | 67 | 56 | 74 | 20 |
| | [Ni] (at%) | / | / | / | 26 |
| | [Mo](at%) | 7 | 18 | / | 4 |
| | Ratio Cr/Fe | 2.6 | 2.2 | 2.8 | / |
| **Passive film 20h** | d (nm) | 1.9 | 0.6 | 1.3 | / |
| | [Fe] (at%) | 26 | 32 | 23 | 53 |
| | [Cr] (at%) | 68 | 49 | 77 | 16 |
| | [Ni] (at%) | / | / | / | 28 |
| | [Mo](at%) | 6 | 19 | / | 2 |
| | Ratio Cr/Fe | 2.6 | 1.5 | 3.3 | / |

Table 3. Best fit parameters for fitting, according to eq. (3) the photo current plots in Fig. 10.

| Sample | $(\mu\tau)_{inn}$ (cm$^2$ V$^{-1}$) | $E_{g\ inn}$ (eV) | $(\mu\tau)_{outer}$ (cm$^2$ V$^{-1}$) | $E_{g\ outer}$ (eV) |
|---|---|---|---|---|
| Native oxide film | $3.0 \times 10^{-18}$ | 3.3 | $4.0 \times 10^{-17}$ | 2.4 |

Table. 4 Best fit parameters of the EIS spectra shown in Fig. 14b.

| Sample | $R_s$ ($\Omega \cdot cm^2$) | $R_P$ ($\Omega \cdot cm^2$) | $Q$ (S·s$^n$cm$^{-2}$) | $n$ |
|---|---|---|---|---|
| Native oxide film | 47.2 | 1.8 10$^5$ | 2.5 10$^{-5}$ | 0.92 |
| Passive film 1 h | 46.4 | 8.3 10$^5$ | 2.2 10$^{-5}$ | 0.92 |
| Passive film 20 h | 43.4 | 8.3 10$^5$ | 2.2 10$^{-5}$ | 0.92 |

**Figure 1**

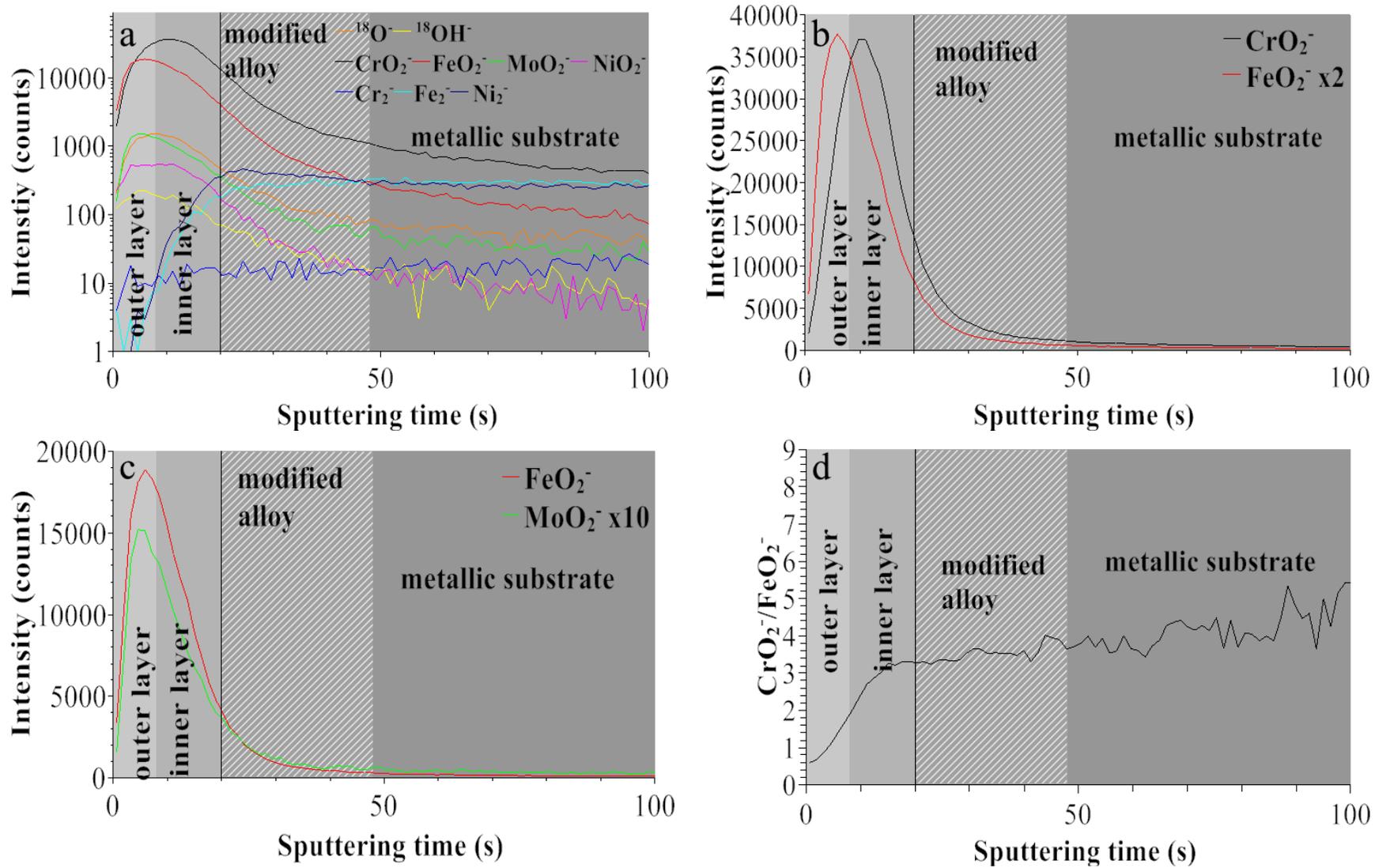

**Figure 2**

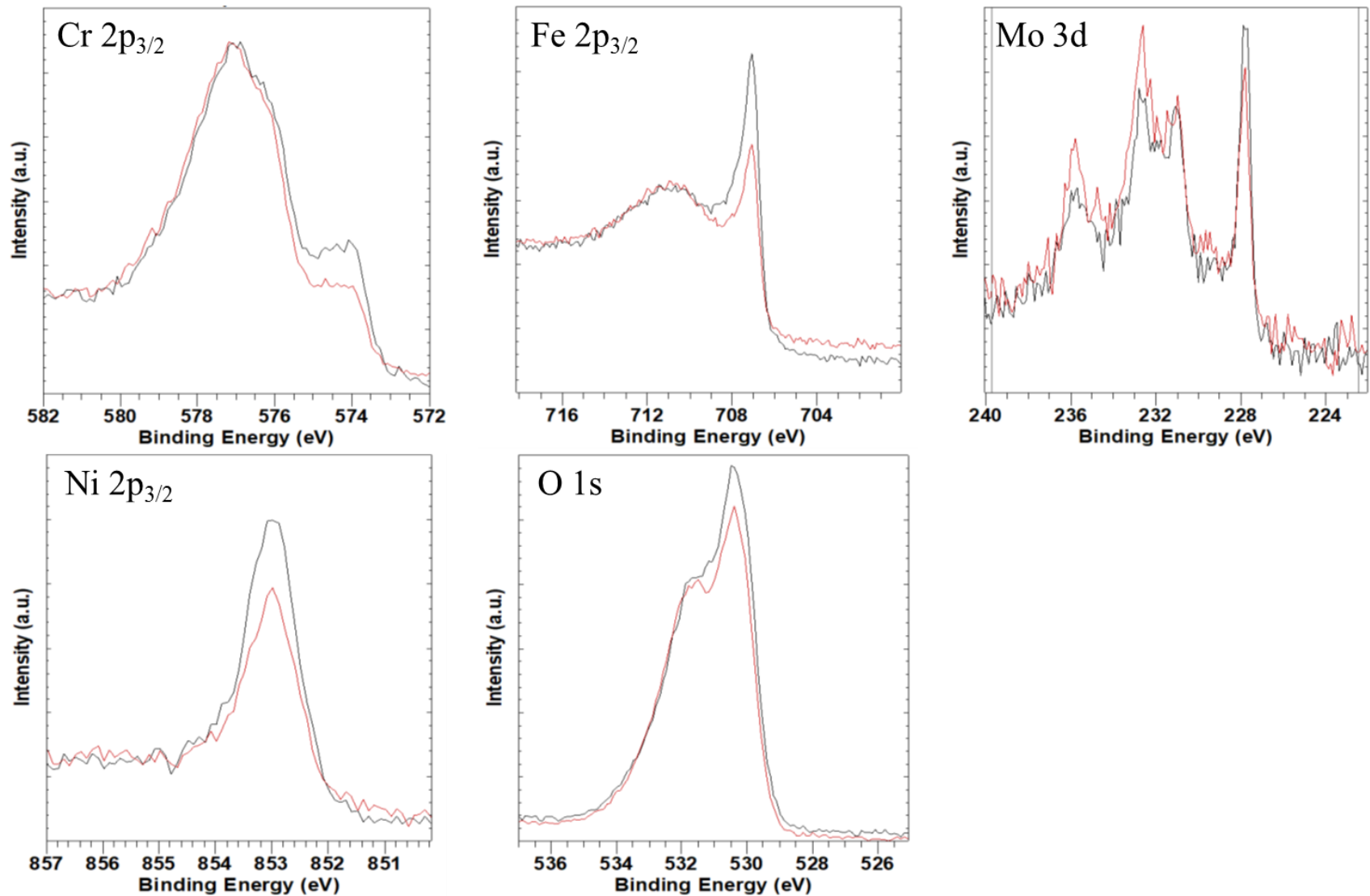

**Figure 3**

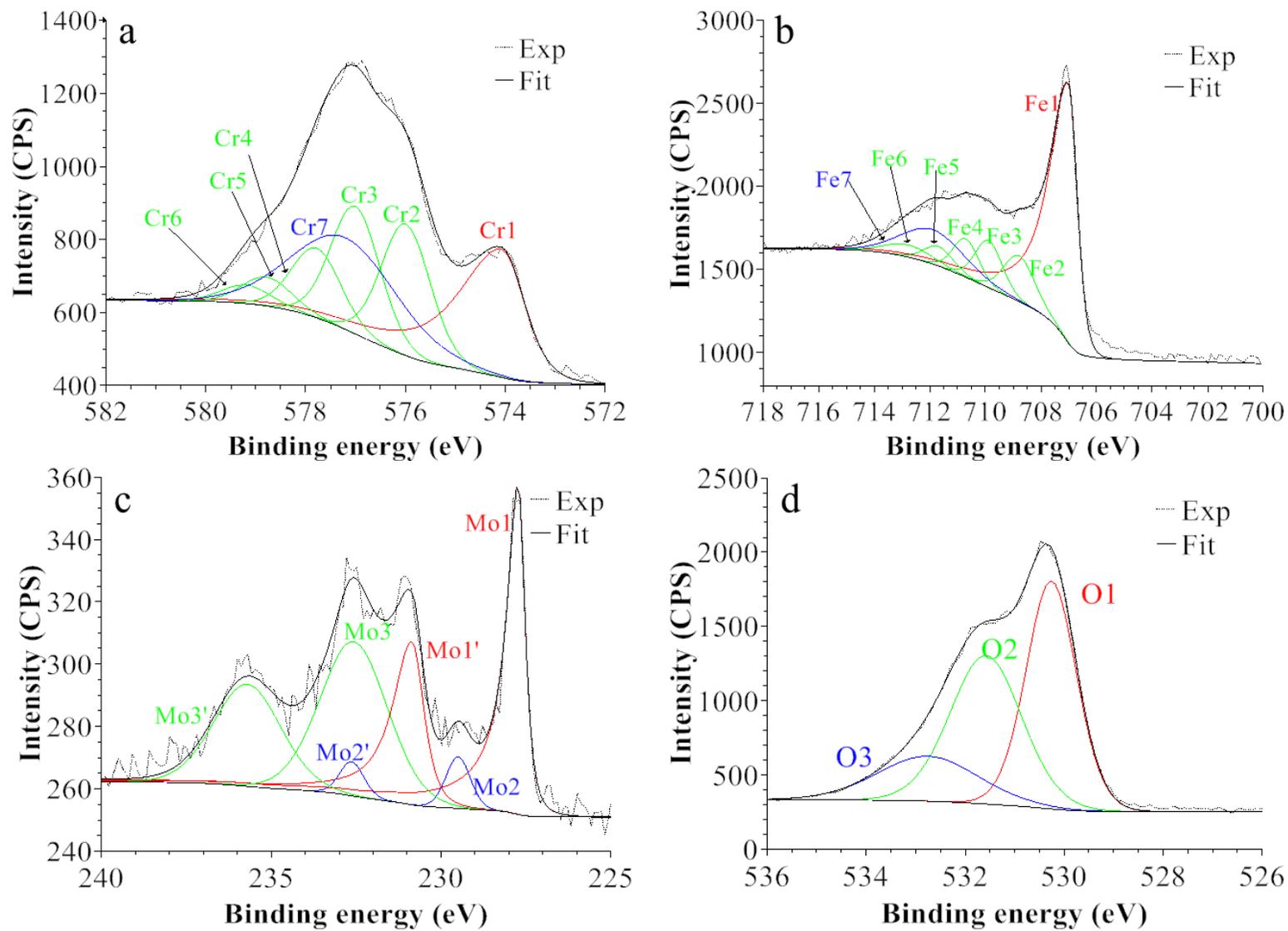

**Figure 4**

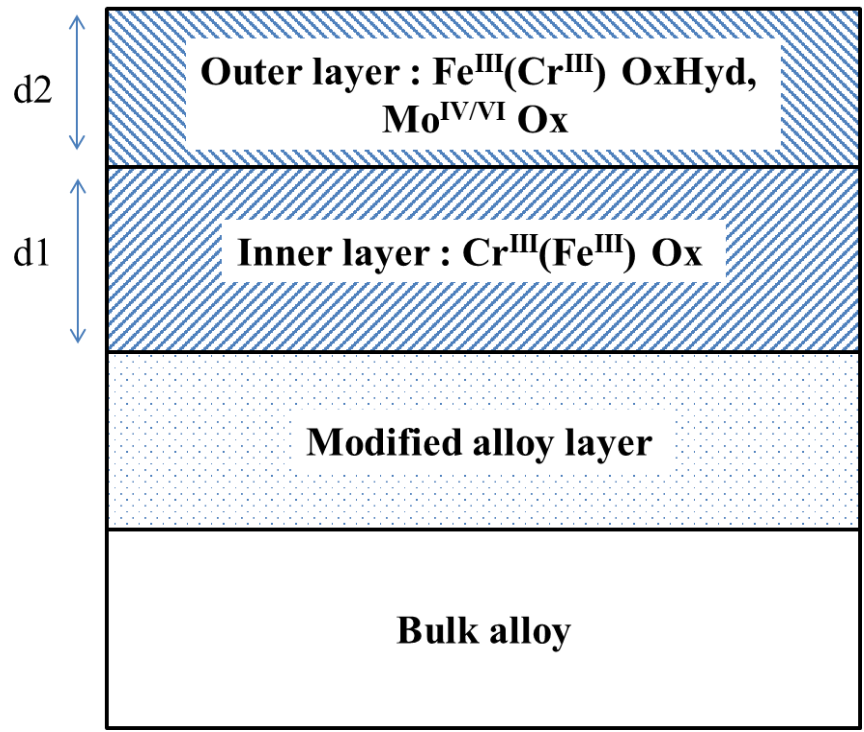

**Figure 5**

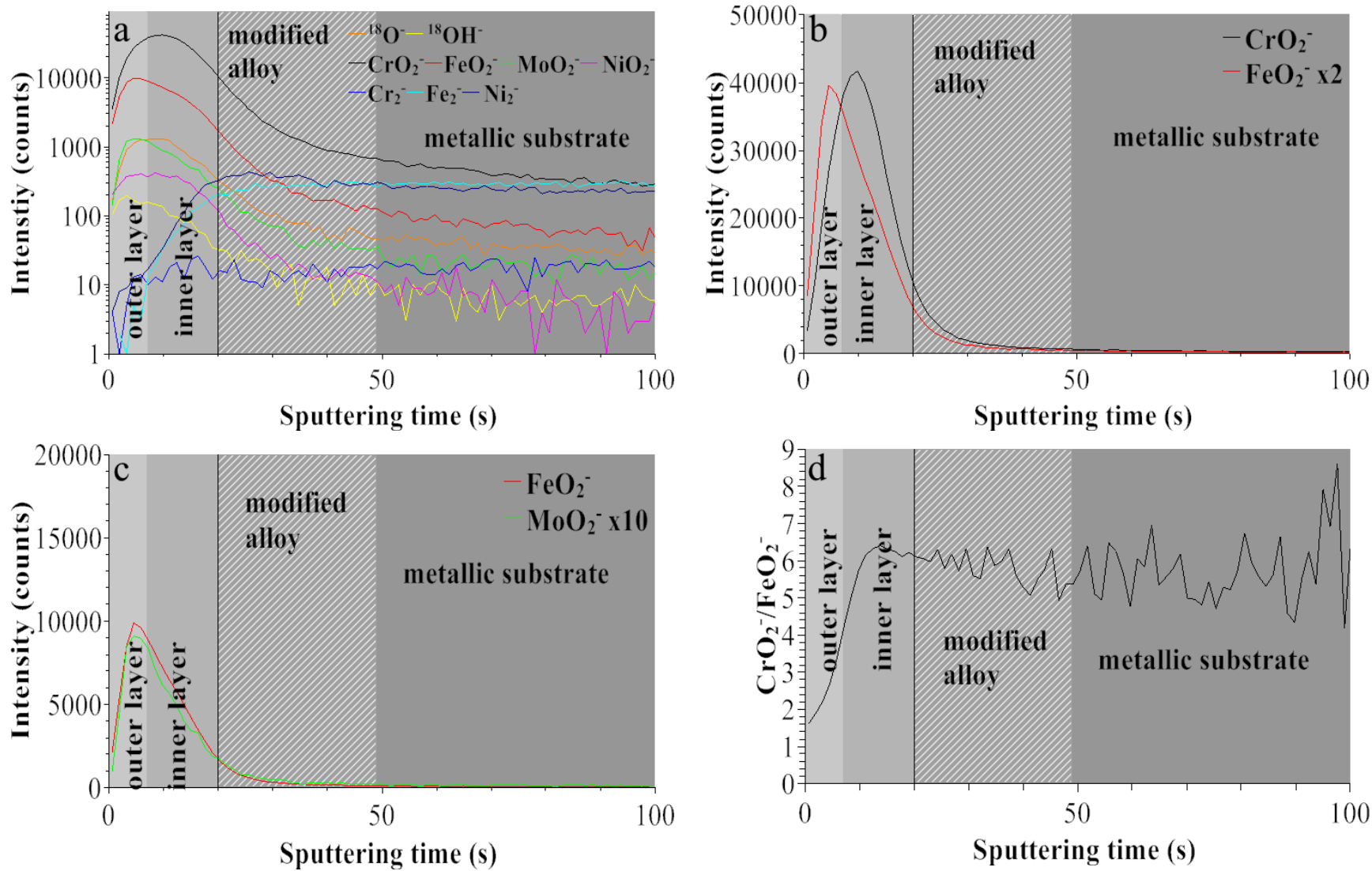

**Figure 6**

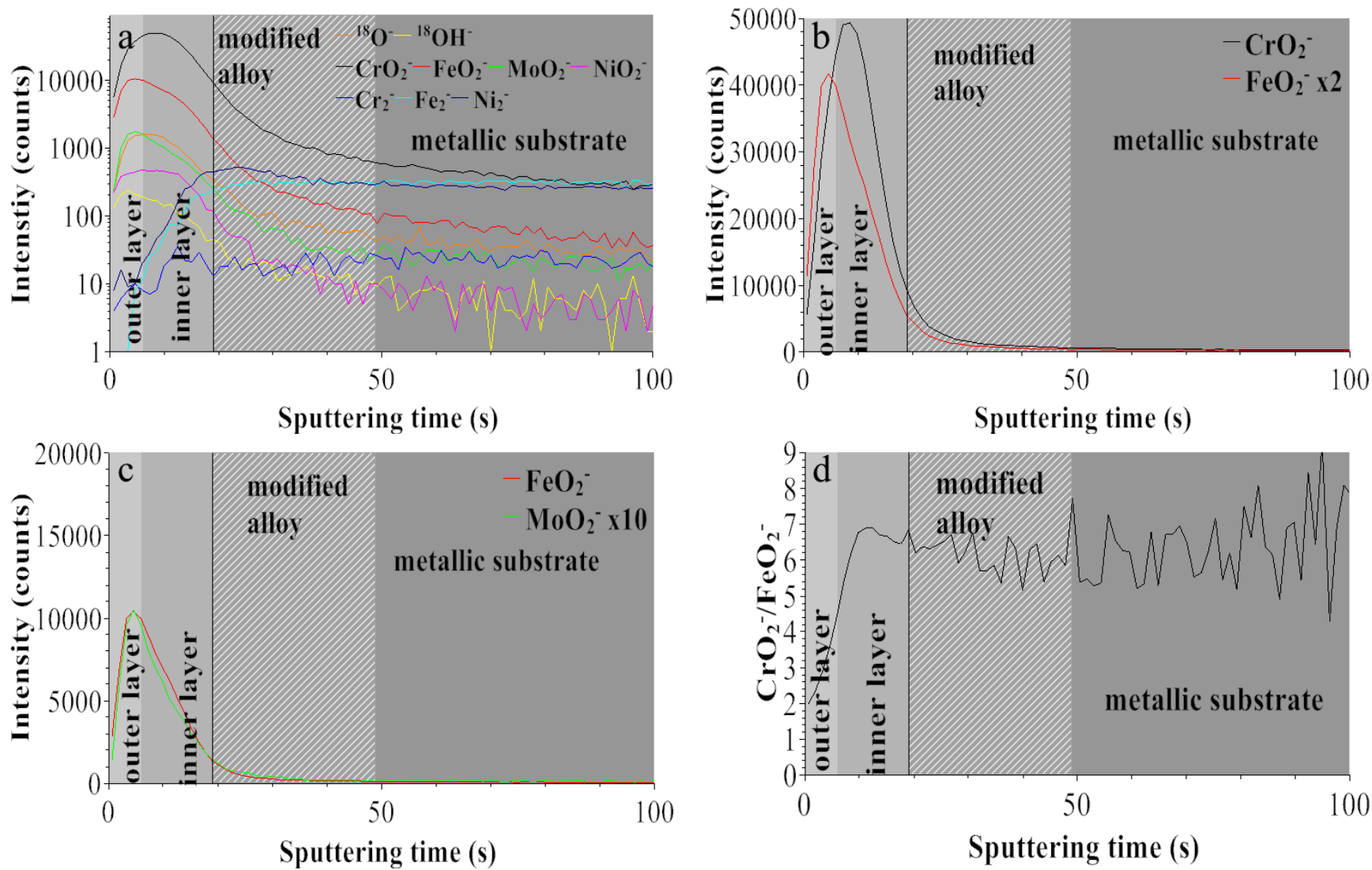

**Figure 7**

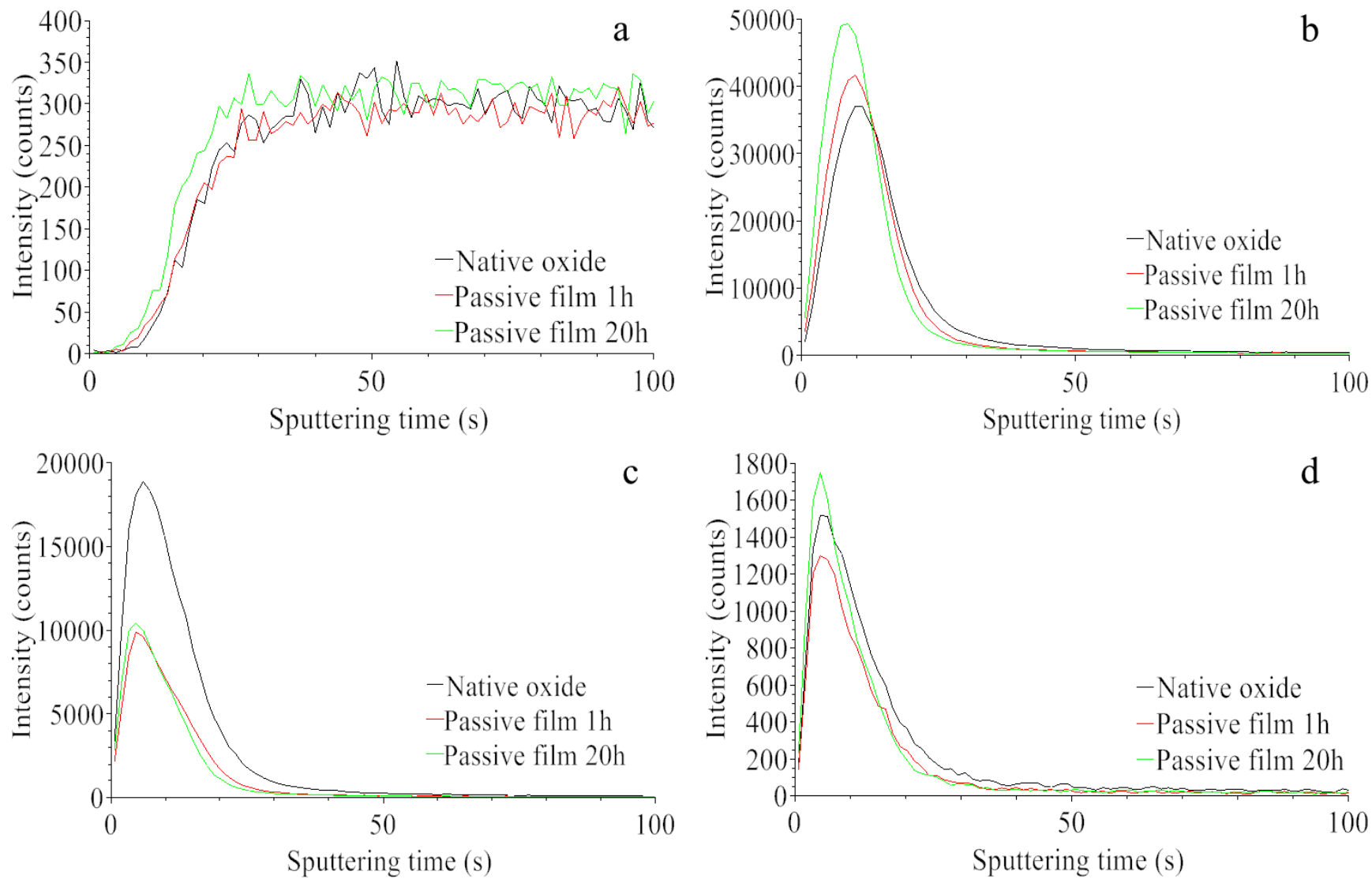

**Figure 8**

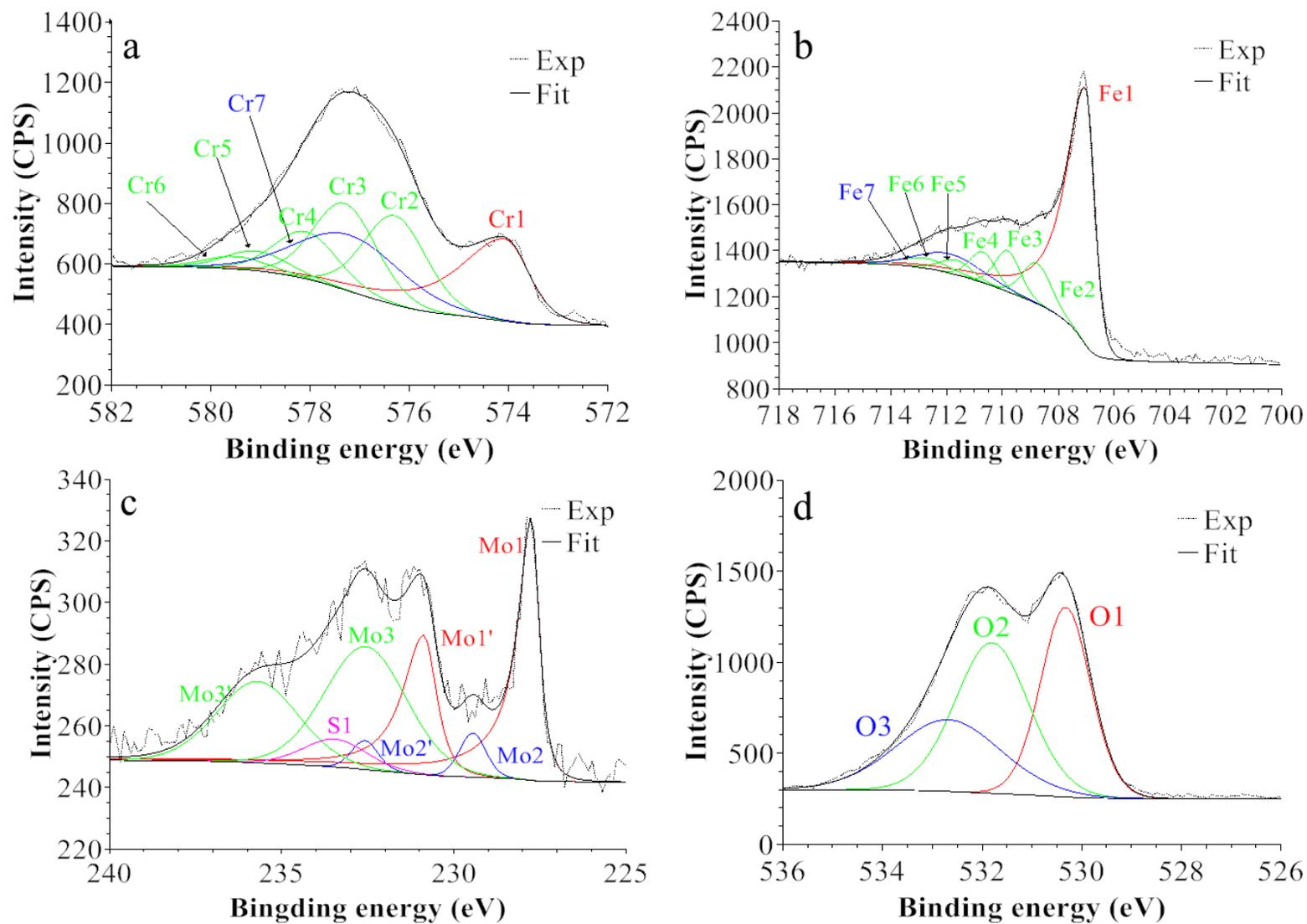

**Figure 9**

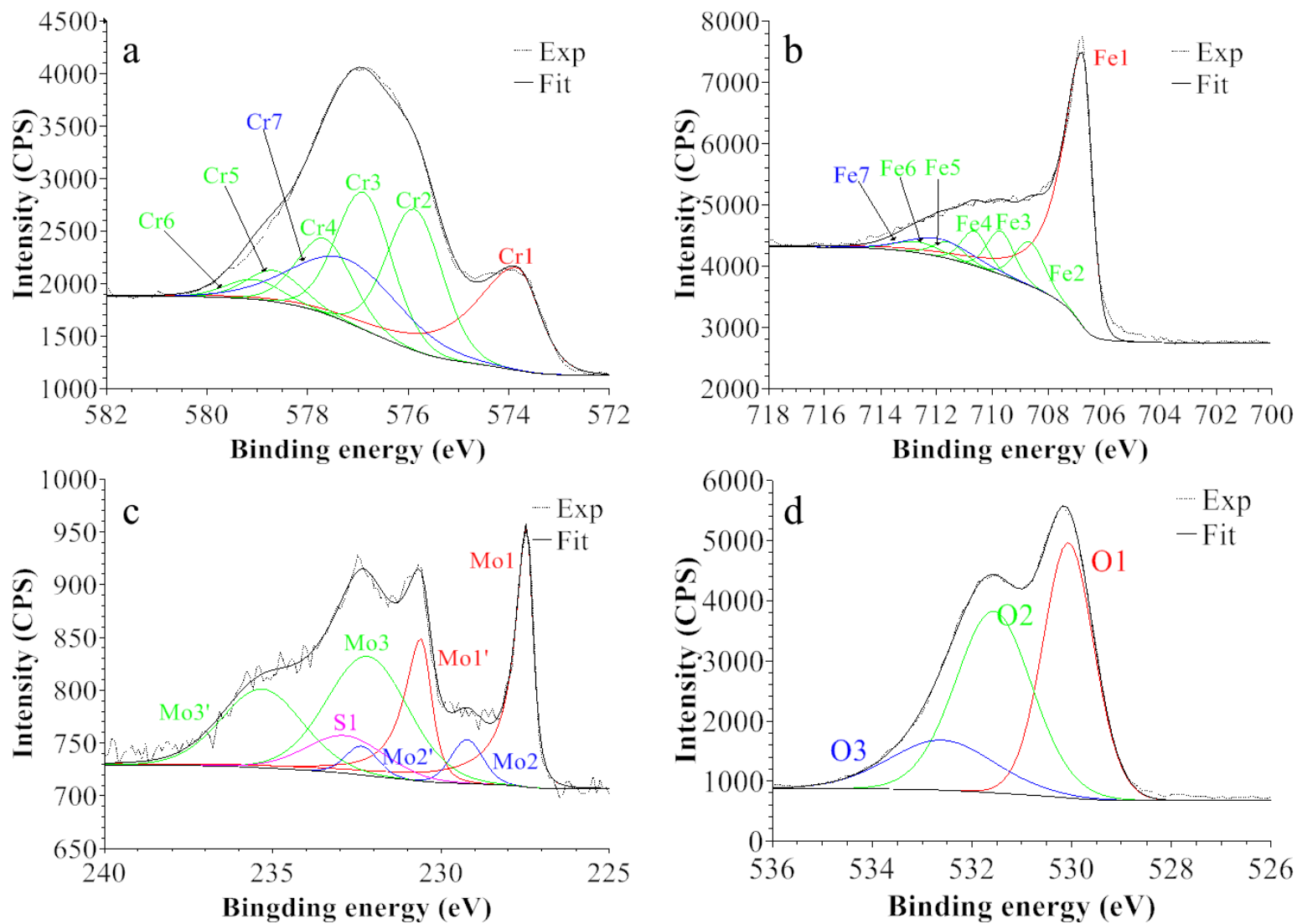

**Figure 10**

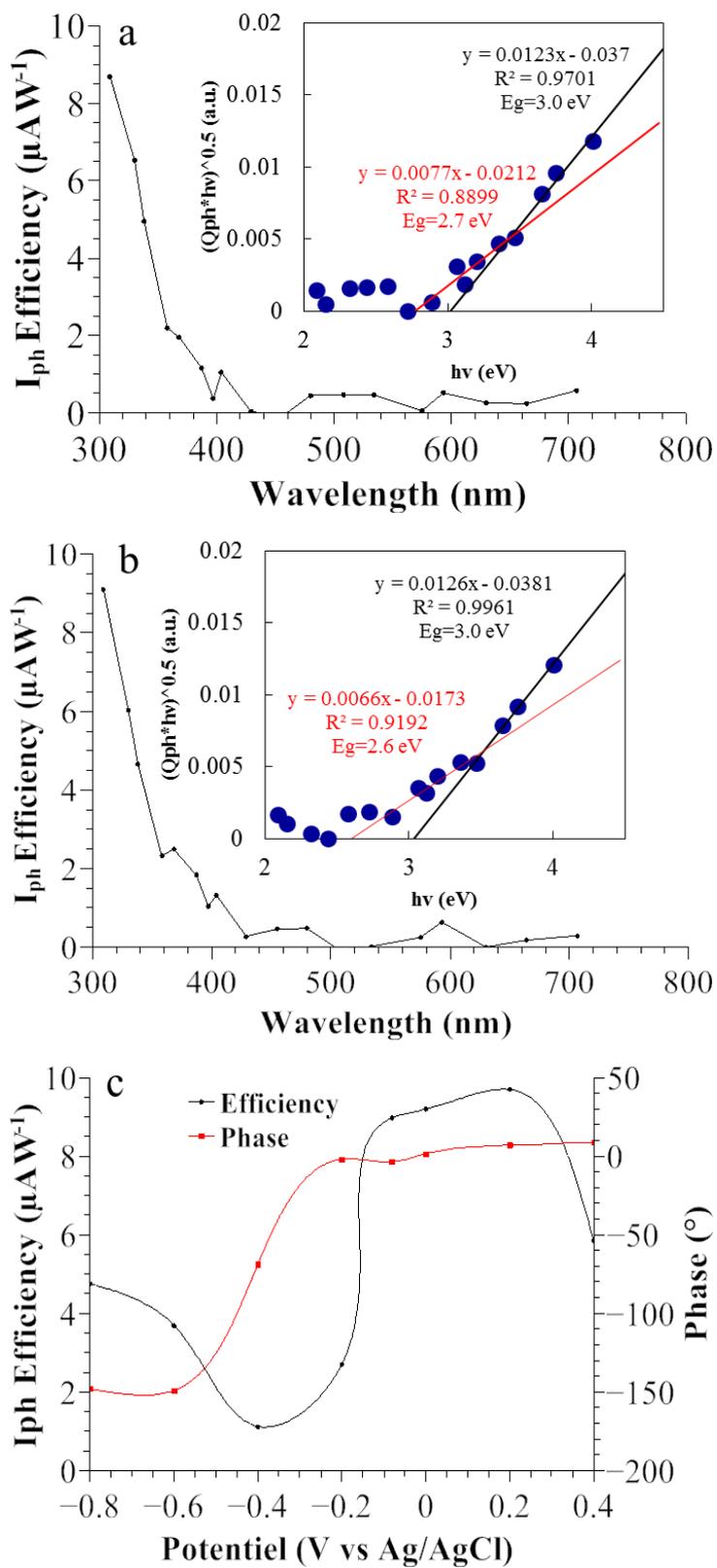

**Figure 11**

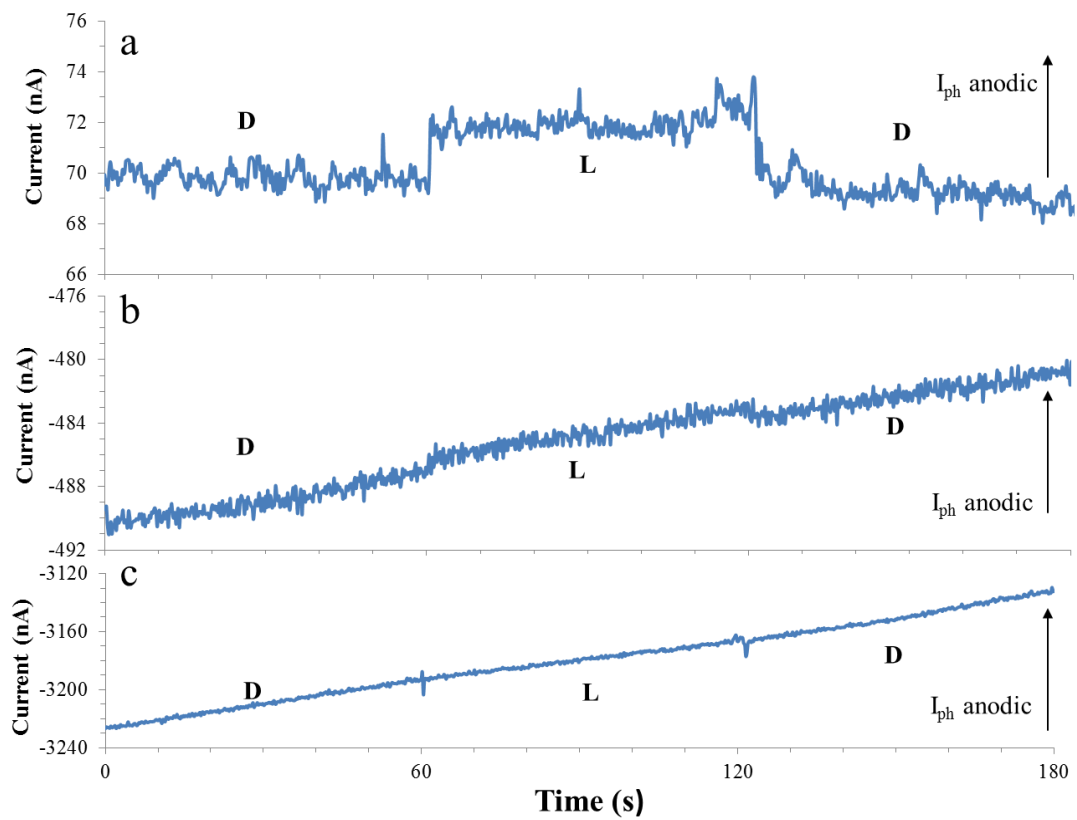

**Figure 12**

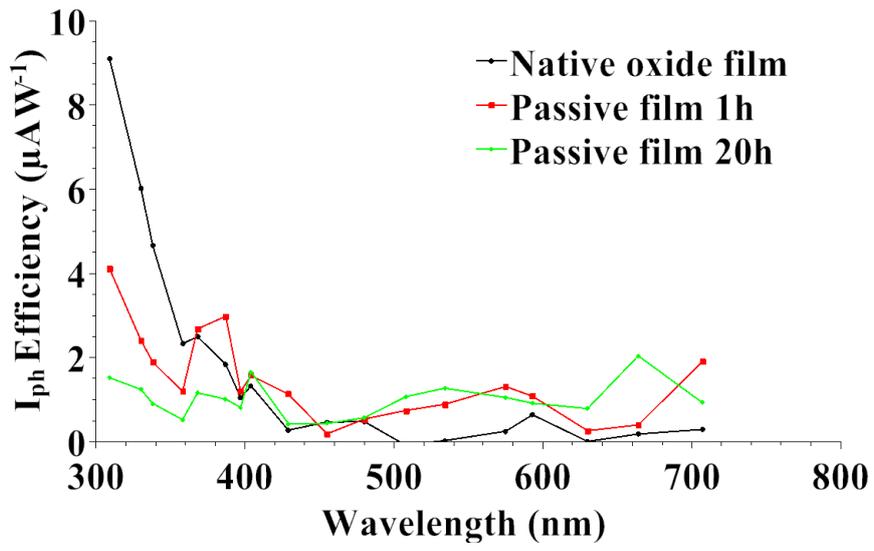

**Figure 13**

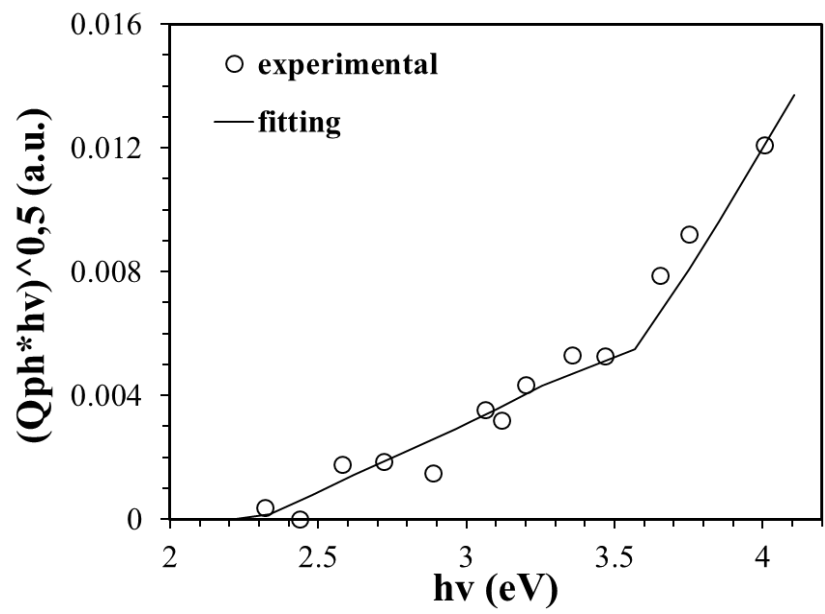

**Figure 14**

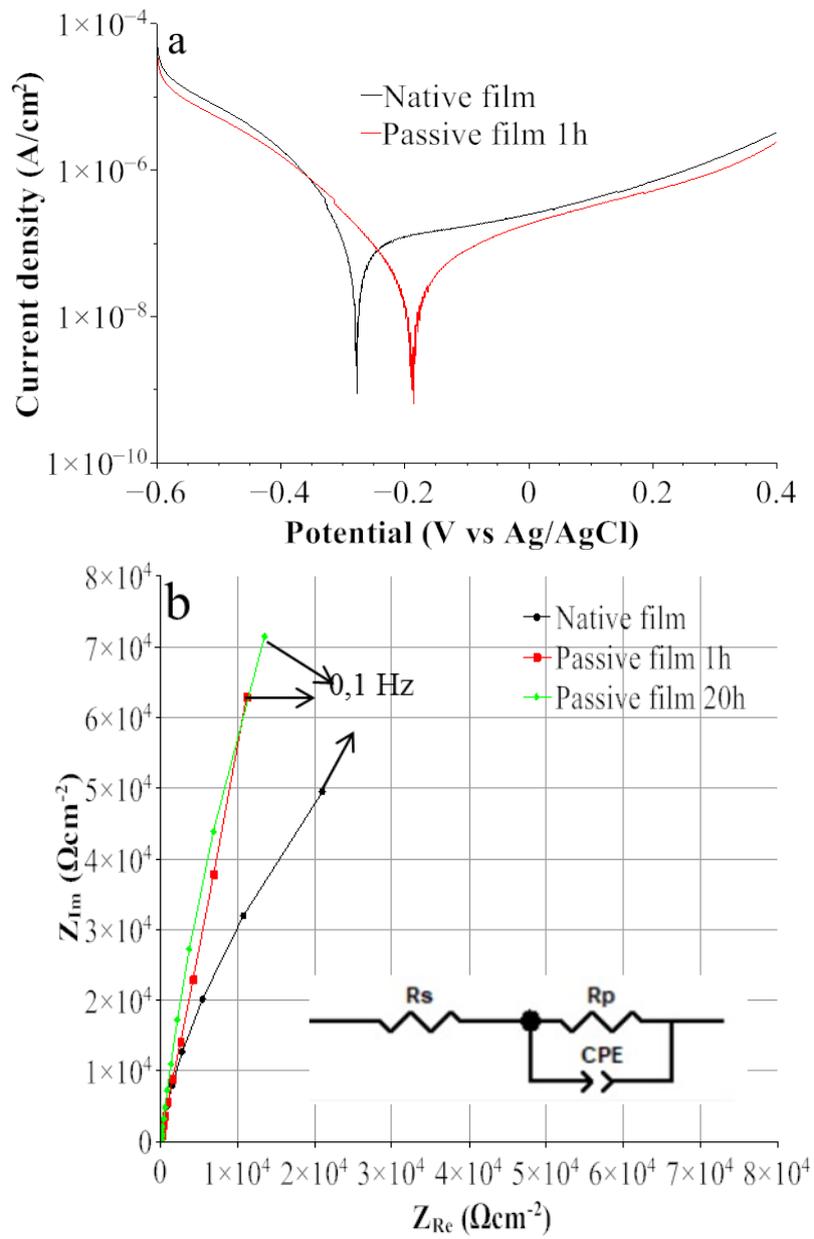